\documentclass[]{spie}  %>>> use for US letter paper
\usepackage[]{graphicx}

\def\3he{$^3{\rm He}$}
\def\4he{$^4{\rm He}$}

\newcommand{\wmap}{{ WMAP}}
\newcommand{\planck}{{\ Planck}}
\newcommand{\herschel}{{ Herschel}}
\newcommand{\blast}{BLAST}
\newcommand{\blastpol}{BLAST-Pol}

\newcommand{\scubatwo}{{ SCUBA-2}}
\newcommand{\scuba}{{ SCUBA}}
\newcommand{\spire}{{ SPIRE}}
\newcommand{\alma}{{ ALMA}}

\def\deg{$^\circ $}
\newcommand{\arcmin}{\hbox{$^\prime$}}               % arcmin
\newcommand{\arcsec}{\hbox{$^{\prime\prime}$}}       % arcsec
\newcommand{\um}{$\mu$m}                            % Microns

\title{The balloon-borne large-aperture submillimeter telescope for polarimetry: BLAST-Pol} 

\author{
       Laura~M.~Fissel\supit{a},
       Peter~A.~R.~Ade\supit{b},
       Francesco~E.~Angil{\`e}\supit{c},
       Steven~J.~Benton\supit{d},
       Edward~L.~Chapin\supit{e},
       Mark~J.~Devlin\supit{c},
       Natalie~N.~Gandilo\supit{a},
       Joshua~O.~Gundersen\supit{f},
       Peter~C.~Hargrave\supit{b},
       David~H.~Hughes\supit{g},
       Jeffrey~Klein\supit{c},
       Andrei~L.~Korotkov\supit{h},
       Galen~Marsden\supit{e},
       Tristan~G.~Matthews\supit{i},
       Lorenzo~Moncelsi\supit{b},
       Tony~K.~Mroczkowski\supit{c},
       C.~Barth~Netterfield\supit{a,d},
       Giles~Novak\supit{i},
       Luca~Olmi\supit{j,k},
       Enzo~Pascale\supit{b},
       Giorgio~Savini\supit{l},
       Douglas~Scott\supit{e},
       Jamil~A.~Shariff\supit{a},
       Juan Diego~Soler\supit{a},
       Nicholas~E.~Thomas\supit{f},
       Matthew~D.~P.~Truch\supit{c},
       Carole~E.~Tucker\supit{b},
       Gregory~S.~Tucker\supit{h},
       Derek~Ward-Thompson\supit{b},
       Donald~V.~Wiebe\supit{e}
\skiplinehalf
\supit{a}{Department of Astronomy \& Astrophysics, University of Toronto, 50 St.~George Street, Toronto, ON M5S~3H4, Canada};
\supit{b}{Department of Physics \& Astronomy, Cardiff University, 5 The Parade, Cardiff, CF24~3AA, UK};\\
\supit{c}{Department of Physics and Astronomy, University of Pennsylvania, 209 South 33rd Street, Philadelphia, PA 19104};\\
\supit{d}{Department of Physics, University of Toronto, 60 St.~George Street, Toronto, ON M5S~1A7, Canada};\\
\supit{e}{Department of Physics \& Astronomy, University of British Columbia, 6224 Agricultural Road, Vancouver, BC V6T~1Z1, Canada};\\
\supit{f}{Department of Physics, University of Miami, 1320 Campo Sano Drive, Coral Gables, FL 33146};\\
\supit{g}{Instituto Nacional de Astrof\'isica \'Optica y Electr\'onica (INAOE), Aptdo. Postal 51 y 72000 Puebla, Mexico};\\
\supit{h}{Department of Physics, Brown University, 182 Hope Street, Providence, RI 02912};\\
\supit{i}{Department of Physics and Astronomy, Northwestern University, 2145 Sheridan Road, Evanston, IL 60208-3112};\\
\supit{j}{University of Puerto Rico, Rio Piedras Campus, Physics Dept., Box 23343, UPR station, San Juan, Puerto Rico};\\
\supit{k}{Osservatorio Astrofisico di Arcetri, INAF, Largo E. Fermi 5, I-50125, Firenze, Italy.};\\
\supit{l}{Department of Physics and Astronomy, University College London, Gower Street, London, WC1E 6BT, UK};
}
\authorinfo{Further author information: (Send correspondence to
  L.~M.~Fissel)\\L.~M.~Fissel: E-mail: fissel@astro.utoronto.ca, Telephone: 416-946-0946}
\begin{document} 
  \maketitle

%%%%%%%%%%%%%%%%%%%%%%%%%%%%%%%%%%%%%%%%%%%%%%%%%%%%%%%%%%%%% 
\begin{abstract}
The Balloon-borne Large Aperture Submillimeter Telescope for Polarimetry (\blastpol) is a suborbital mapping experiment designed to study the role played by magnetic fields in the star formation process. \blastpol\ is the reconstructed \blast\ telescope, with the addition of linear polarization capability.
Using a 1.8\,m Cassegrain telescope, \blastpol\ images the sky onto a focal plane that consists of 280 bolometric detectors in three arrays, observing simultaneously at 250, 350, and 500\,$\mu$m. The diffraction-limited optical system provides a resolution of 30\arcsec at 250\,\um.
The polarimeter consists of photolithographic polarizing grids mounted in front of each bolometer/detector array. A rotating 4\,K achromatic half-wave plate provides additional polarization modulation.
With its unprecedented mapping speed and resolution, \blastpol\ will produce three-color polarization maps for a large number of molecular clouds. The instrument provides a much needed bridge in spatial coverage between larger-scale, coarse resolution surveys and narrow field of view, and high resolution observations of substructure within molecular cloud cores. The first science flight will be from McMurdo Station, Antarctica in December 2010.
\end{abstract}

%>>>> Include a list of keywords after the abstract 


%%%%%%%%%%%%%%%%%%%%%%%%%%%%%%%%%%%%%%%%%%%%%%%%%%%%%%%%%%%%%
\section{INTRODUCTION}
\label{sec:intro}  % \label{} allows reference to this section
\blastpol, the Balloon-borne Large Aperture Submillimeter Telescope for Polarimetry, is a
stratospheric 1.8\,m telescope which maps linearly polarized submillimeter emission with bolometric
detectors operating in three 30\%\ wide bands at 250, 350, and
500$\,\mu$m. \blastpol's diffraction-limited optics were designed to provide a resolution of 30\arcsec, 42\arcsec, and 60\arcsec\ at the
three wavebands, respectively. The detectors and cold optics are
adapted from those used by the SPIRE instrument on \herschel\
\cite{grif03}.

\blastpol\ is a rebuilt and enhanced version of the \blast\ telescope\cite{pascale08}, with added linear polarization capability.  \blast\
 was designed to conduct confusion-limited, wide-area
extragalactic and Galactic surveys at submillimeter (submm)
wavelengths from an balloon platform.  

\blast\ had two Long Duration Balloon (LDB) flights.  The first was a 4-day flight from Kiruna, Sweden in June of 2005 (BLAST05).  Unfortunately the telescope was found to be out of focus, due to slight damage of the primary mirror during the launch or ascent, so the telescope was restricted to observing bright Galactic targets.  \blast\ was repaired and flown again from Antarctica in December 2006 for 11 days (BLAST06).  For BLAST06 the instrument worked perfectly and multiple deep, large-area maps were obtained for Galactic and extra-galactic fields. 
 
After termination of the Antarctic flight the parachute could not be detached, which resulted in the payload being dragged $\sim$200\,km along the ground in 24\,hours. \blast\ was largely destroyed, but fortunately the pressure vessel containing the hard drives which stored all of the experiment data was recovered.  In addition, the mirrors, detectors and receiver were all recovered, and have been used in the construction of \blastpol.

The \blast\ telescope has left a legacy of fantastic science results.  \blast\ provided the first deep, wide area maps at 250, 350 and 500 \um, bands that are very difficult or often impossible to observe from even the best ground based sites in the world.  Some science highlights include measurement of the the FIR background at 250, 350 and 500 \um, including a 0.8 deg$^2$ confusion-limited map in the GOODS South region, where it was shown that more than half of the FIR background light originates in galaxies at redshift$>$1.2 \cite{devlin09,marsden09,pascale09}, the first determination of deep, extragalactic number counts at these wavelengths\cite{patanchon09}, the detection of clustering in the far-IR background \cite{viero09}, three-band resolved sub-mm images of several nearby galaxies\cite{wiebe09}, and the determination of luminosities, masses and temperatures of more than a thousand compact sources in the Vela Molecular cloud, which may form into stars \cite{netterfield09}.

With the addition of a polarimeter, \blast\ has now been 
transformed into \blastpol, a uniquely sensitive polarimeter 
for probing linearly polarized Galactic dust emission. 
\blastpol\ will map the magnetic field structure of many 
different star-forming regions. These maps will provide a 
fantastic data set for addressing what role magnetic fields play in the star formation process, an important outstanding question in our understanding of how stars form. \blastpol\ will be able to map magnetic fields across entire Giant Molecular Clouds (GMCs), with sufficient resolution to probe fields in dense filamentary sub-structures and molecular cores. The experiment provides a crucial bridge between the large area but coarse resolution polarimetry provided by experiments such as \planck\ (5\arcmin\ resolution) with the resolution of the \alma\ telescope.

\blastpol's first flight is scheduled for December 2010 from McMurdo Station, Antarctica.

\section{SCIENCE GOALS}
\label{sec:sci}  % \label{} allows reference to this section

%%%%%%%%%%%%%%%%%%%%%%%%%%%%%%%%%%%%%%%%%%%%%%%%%%%%%%%%%%%%%
\subsection{Probing the Role of Magnetic Fields in Star Formation}
Many fundamental questions about the process of star formation
 remain unanswered\cite{mckee07}.  Is star formation regulated by
magnetic fields or by turbulence? How long does the star formation 
process last? Do molecular clouds and their associated substructures 
(dense cores, filaments, and clumps), have lifetimes exceeding
their turbulent crossing times?  What determines the final masses of 
stars?

It is difficult to observe magnetic fields in molecular 
clouds\cite{crutch04,whittet08}.  One promising method for 
probing them is to observe clouds with a far-IR/submm
polarimeter\cite{hildebrand00, ward00}.  By tracing the linearly polarized 
thermal emission from dust grains aligned with respect to the 
local magnetic fields, we can measure the direction of the
plane-of-the-sky component of the magnetic field within the cloud.  
By mapping polarization we can therefore trace the magnetic 
field direction across entire molecular clouds.  

\begin{figure}
  \begin{center}
  \begin{tabular}{ccc}
    \includegraphics[height=5.8cm, keepaspectratio]{f1.eps} && \includegraphics[height=5.8cm, keepaspectratio]{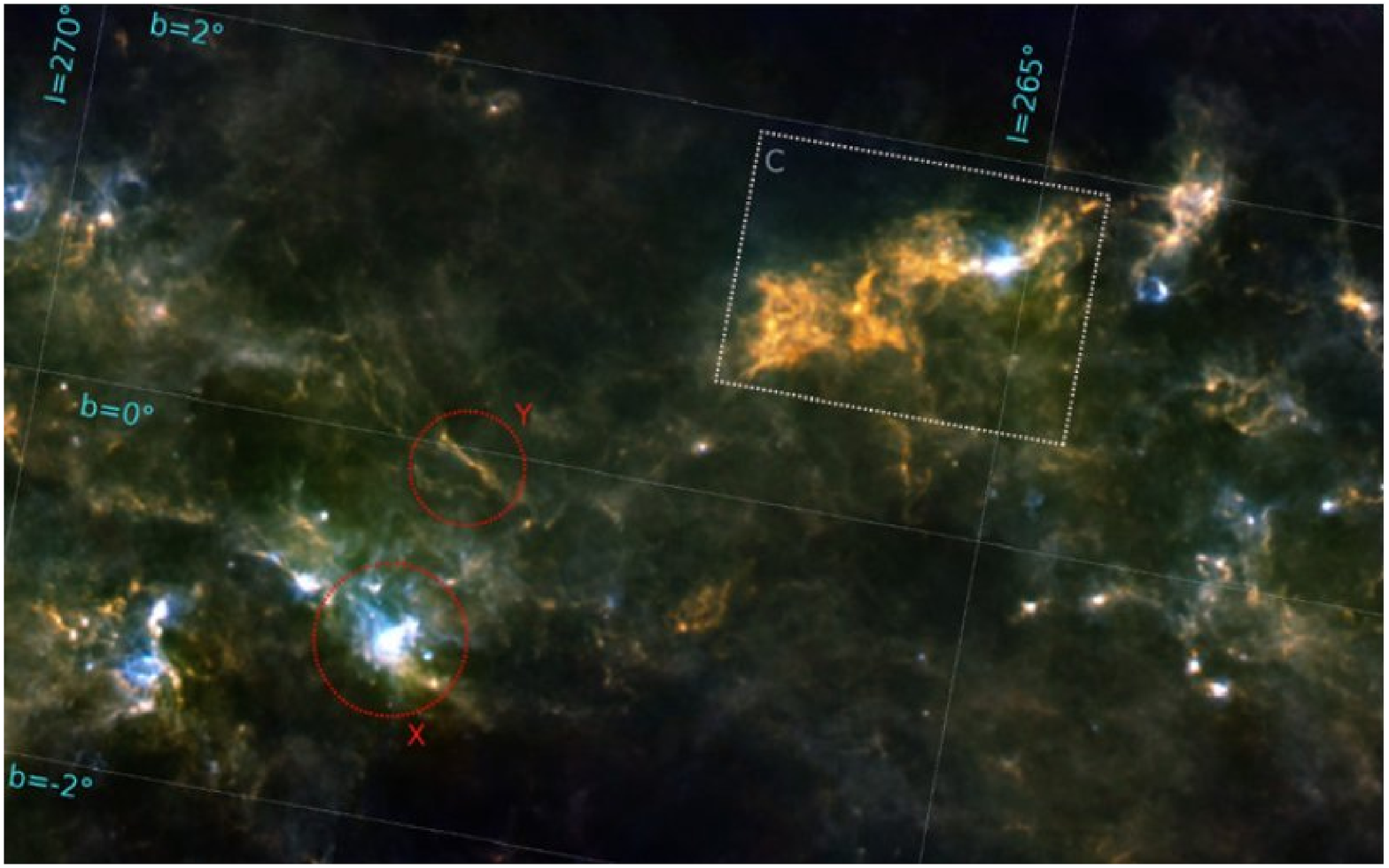}
  \end{tabular}
  \end{center}
  \caption[example] { \label{fig:mag1} {\em Left:\/}
    Submillimeter polarimetry (at 450\,\um) of NGC 6334, a 
    massive GMC at a distance of 1700 pc. The pseudo-vectors 
    in the map at upper left were obtained using the SPARO 
    instrument at the South Pole, and the ``blowup'' images at the right
    and bottom are from Hertz/CSO observations on Mauna Kea 
    at 350 \um\cite{novak09}. The orientation of
    each pseudo-vector shows the inferred field direction, with the length
    of each line proportional to the degree of
    polarization. {\em Right:} A excerpt from a 50 deg$^2$\ false color BLAST06 
    map of the Vela Molecular Cloud complex\cite{netterfield09}.  
    Vela C, a GMC located at a distance of 
    $\sim$700\,pc and a planned \blastpol\ target, is 
    indicated with a dashed rectangle.
    During the first \blastpol\ flight, we will map $\sim$20 such clouds,
    obtaining $\sim$1000 polarization pseudo-vectors per cloud.  }
\end{figure}

Far-IR/submm polarimetry is an emerging area of star formation research,
 with many
upcoming experiments that will map fields on different scales. 
\planck\cite{lamarre03} will provide coarse resolution 
(FWHM\,$\sim 5\arcmin$) submm polarimetry maps of the entire sky.  
\alma\ will provide sub-arcsecond resolution polarimetry 
capable of resolving fields within cores and disks, but will not be 
sensitive to cloud-scale fields.  \blastpol, with a 30\arcsec\
 FWHM beam at 250\,\um, will be the first submm polarimeter 
to have both the sensitivity and mapping 
speed necessary to trace fields across entire clouds for 
a statistically significant sample of molecular clouds, and 
sufficient spatial resolution to track the cloud fields into 
the dense substructures and cores within the clouds. 
Observations from the SPARO experiment show that 
extended submm emission from giant molecular clouds (GMCs), 
where most stars are thought to have formed, 
is indeed polarized\cite{li06,novak09} (Fig.~\ref{fig:mag1}). \blastpol\ 
will map polarized emission over a wide range of dust column 
densities corresponding to $A_v$\,$\ge$4\,, yielding
 $\sim 1000$ independent polarization detections per cloud, for 
dozens of clouds.

The \blastpol\ maps will allow us to make the first detailed comparisons 
between observed molecular cloud field maps and synthetic maps.  
The latter can be derived from numerical simulations of 
molecular clouds (for example see Ref. \citenum{ostriker01}). 
This will allow us to make detailed
observational tests of theoretical magneto-hydrodynamic (MHD)
models of star formation. 

In particular, we will address the following questions with \blastpol:

\begin{enumerate}
\item {\em Is molecular core morphology determined by large-scale magnetic fields?} Jones \&
  Basu\cite{jones02} have argued that observations show a predominance
  of oblate-shaped cores in molecular clouds, as has been predicted by
  magnetically-regulated models of core collapse\cite{mous99, allen03}. 
  These models also predict that each core should be embedded in a 
  large-scale cloud field which runs parallel to the core's minor axis.  
  Submm polarimetry observations of 4 quiescent cores with \scuba \cite{ward00} 
  questioned this prediction by finding that the magnetic fields were 
  misaligned, and therefore more consistent with turbulence dominated magnetic 
  models. More recently Tassis et al.\cite{tassis09} analyzed observations of 24 clouds taken with the 
  Hertz polarimeter at the Caltech Submillimeter Observatory (CSO) and 
  found in their statistical analysis that models with oblate cores and mean 
  magnetic field orientations with small deviations from the core 
  minor axis are preferred.  However, observations of more clouds 
  are needed in order to reject models.
  Also, Hertz, \scuba\ (and \scubatwo) cannot realistically
  detect fields in lower density regions: what is
  needed are observations tracing core fields out into the surrounding
  lower-density cloud environment.  \blastpol\ will provide these for a large 
  sample of molecular clouds.
\item {\em Do filamentary structures within clouds have magnetic origins?}  
  Faint, low column density $^{12}$CO filaments observed in
  Taurus\cite{heyer08,goldsmith08} seem to closely follow magnetic 
  field lines traced by optical polarimetry, which could indicate 
  streaming of molecular gas along field lines.
  However, denser filaments in Taurus and some other clouds show no
  preferred orientation with respect to the field direction traced 
  by nearby polarimetry pseudo-vectors from optical observations of background 
  stars\cite{goodman90}. This may imply a non-magnetic origin for
  the denser filaments, or it may simply reflect the inadequacies of
  optical/near-IR polarimetry for tracing fields in dense, shielded
  regions within molecular clouds\cite{whittet08,cho05}.  In contrast, 
  optical polarization observations of the Pipe Nebula
  \cite{alves08} show that the local magnetic field is perpendicular to 
  the filamentary structure.  We will use \blastpol\ to map
  dense filaments to answer questions about
  the relationship between fields and cloud sub-structures. 
\item {\em How strong are magnetic fields in molecular clouds, 
  and how does the field strength vary from cloud to cloud?}  
  Simulations have shown that clouds where magnetic fields 
  are strong enough to play an important role in supporting 
  the cloud against gravitational collapse tend to have  
  aligned polarization angles, where as clouds with 
  more randomly oriented polarization angles imply weaker fields\cite{ostriker01}. 
  Field strength can be estimated using the Chandrasekhar-Fermi 
  (CF) technique\cite{chandra53}, which relates 
  dispersion in polarization angle to field strength\cite{chandra53,zweibel90}.   CF field strength estimates for molecular cloud cores have been
  obtained from submm data, and the derived magnetic field strengths are in 
  rough agreement those derived from Zeeman observations\cite{crutch04}.  
  Numerical turbulence simulations have been used to calibrate 
  the CF technique for molecular clouds\cite{ostriker01,pelkonen07,
  falceta08}. \blastpol\ will determine the power spectrum of the 
  polarization angle dispersion over a large range of spatial scales, 
  for dozens of molecular clouds. These data 
  should lead to improved models and better CF calibration.
  \blastpol\ will also measure 3-color polarization spectra 
  variations within the cloud \cite{bethell07,vaillancourt08}, which 
  will allow us to explore the dust properties and temperature 
  structure of each cloud.  Our observations will 
  probe the dependence of field strength on cloud age,  
  location, and mass.
\end{enumerate}

\subsection{Polarized Galactic Dust Foregrounds}
The search for primordial gravitational waves through detecting the
B-mode polarization of the Cosmic Microwave Background
(CMB)\cite{hu97,kamionkowski97,zaldarriaga97} requires precise
control of instrumental and astrophysical systematics at a level which
allows the detection of this extremely faint signal. 
\wmap\ polarization observations\cite{page07} clearly demonstrate that
polarized foregrounds caused by diffuse Galactic dust emission are likely to be comparable or
larger than the B-mode CMB polarization anisotropy at large angular
scales (multipoles  $\ell < 100$) and at intermediate Galactic latitudes. This
situation differs from the CMB temperature anisotropy signal where 
most of the sky is dominated by the CMB.  It requires significant 
foreground emission
modeling in order to extract the underlying CMB polarization
signal. 

Most of the polarized Galactic foregrounds observed by \wmap\ are 
due to synchrotron emission, but at the highest \wmap\ frequency band
(94\,GHz), there is evidence for polarized dust emission. This 
polarized dust emission is
potentially problematic for many higher-frequency bolometric CMB
polarization experiments (for example EBEX, Spider, Clover, QUAD, BICEP, Keck,
\planck-HFI, and Polarbear), since the emission is expected to rise
significantly with frequency as predicted by models based on
observations at 100$\,\mu$m. The ARCHEOPS
balloon mission measured Galactic dust polarization on scales of
15\arcmin\ at submm wavelengths. ARCHEOPS results\cite{ben04} have
shown that radiation from diffuse Galactic dust is polarized (as
expected) at the $3-5$\%\ level, but in some regions the polarization is
as high as 10\%. While these measurements were made near the Galactic
plane, they point to the need for more information at higher
latitudes, with higher sensitivity, better angular resolution, and at
multiple wavelengths. Despite the \wmap\ and Archeops results, no
information exists for any region of the sky at the accuracy required
for a B-mode signal detection, and very little information exists at
frequencies relevant to CMB science (at least until \planck\ results 
are published). 

There are additional puzzles in dust
polarimetry that must be resolved if anyone is to make a robust detection
of the CMB B-modes. For example, polarimetry of bright molecular
cloud cores at 350$\,\mu$m and 1\,mm has revealed the surprising result
that the polarization fraction at 1\,mm is larger by a factor of $\sim$2
compared to that at 350$\,\mu$m\cite{hildebrand00}. The
\planck\ satellite will provide a unique data set by mapping the sky
with polarization-sensitive bolometers in bands centered from
850$\,\mu$m to 3\,mm. The 850$\,\mu$m band will provide an
unparalleled data set for studying polarized dust emission. In addition to
\planck, balloon-borne and other ground-based instruments employ
anywhere between 2 to 6 frequency bands to assist in discriminating against
foreground contamination. But the ability to separate the foreground
dust emission to the necessary level to detect B-mode polarization
depends on the complexity of the polarized dust emission.

\begin{figure}
  \begin{center}
  \begin{tabular}{c}
    \includegraphics[width=0.4\linewidth, angle=-90,keepaspectratio]{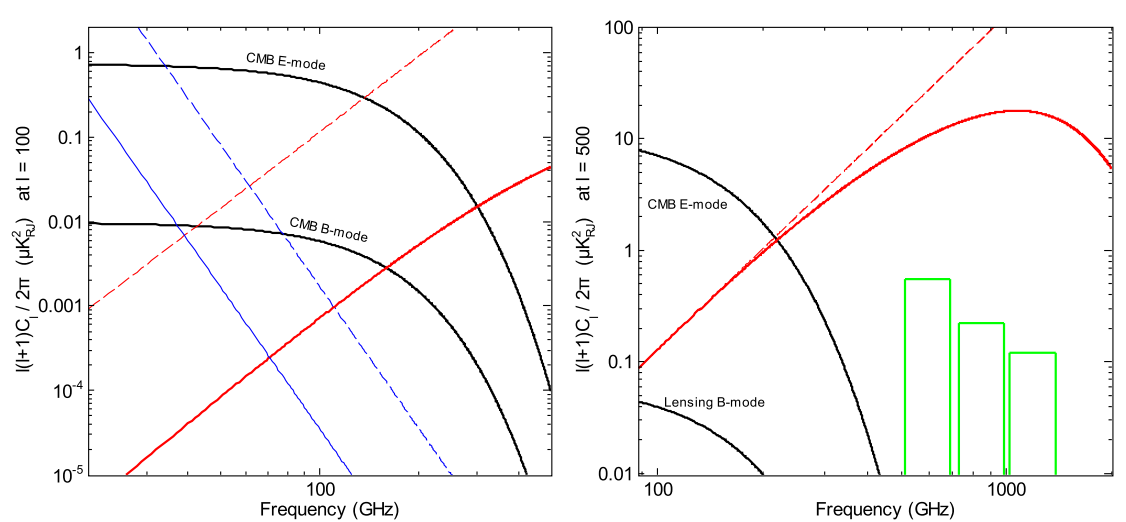}
  \end{tabular}
  \end{center}
  \caption[example] { \label{fig:foregrounds} {\it Left:\/} The
    spectrum of B-mode foregrounds from Galactic dust (red) and
    synchrotron (blue) expected outside \wmap's P06 mask\cite{page07}
    (dashed), and from a selected sky region (solid), compared with the E and
    B-mode CMB (black, solid). These are all at $\ell = 100$ and in antenna
    temperature units. This dust model assumes model 8 from Finkbeiner et
    al. (1999)\cite{fink99}. {\it Right:\/} A similar plot at $\ell
    =500$. The green bars show the \blastpol\ noise at a resolution of
    8\arcmin\ from a 48-hr observation of 0.5 deg$^2$. At these
    angular scales, the CMB signal is dominated by the lensing B-modes. The
    Galactic dust signal, normalized to \wmap, peaks at the
    \blastpol\ frequencies. (The CMB models here assume a standard
    $\Lambda$CDM cosmology with {\em r} = 0.1.)  }
   \end{figure}

Figure~\ref{fig:foregrounds} shows the expected polarized dust
emission as measured by the 3\,yr \wmap\ data\cite{page07} and the level
of foregrounds predicted in a selected $\sim 200$\,deg$^2$ sky region
($l = 258$\deg, $b = -46$\deg) accessible by sub-orbital and ground-based 
instruments.  Amazingly, the dust signal extrapolated to the CMB
frequencies is comparable to the gravitational-wave contribution {\em r}\,=\,0.1 B-mode signal level that
is being probed by the current generation of experiments. For this
reason, it is imperative that we begin to understand high-latitude
polarized dust emission. Given these rather low signal levels,
\blastpol\ will be limited in its ability to conduct large surveys that could be
used as templates for lower frequency CMB polarization observations.
However, a 48-hr survey over a 0.5 deg$^2$ sky region
(Fig.~~\ref{fig:foregrounds}, right panel) will enable \blastpol\ to
constrain dust properties which are difficult to measure at lower
frequencies. In particular, the determination of dust temperature and
 degree of polarization will be highly degenerate as measured by
all CMB experiments in the Rayleigh-Jeans tail of the dust spectrum.
On the other hand, the ability of \blastpol\ to extract these parameters will be
straightforward, since the measurements are made near the peak of the cold dust
spectrum.

The first of our two maps will be aimed at a very low dust emission
region, comparable to 1 MJy\,sr$^{-1}$ at 100$\,\mu$m, that is ideally
coincident with a deep CMB polarization measurement. The other will
occur in a mid-latitude region with a higher dust brightness that will
almost certainly be sampled by \planck. The corresponding higher signal
will enable parameter extraction without the aid of lower frequency
CMB polarization measurements.

When these measurements are combined, they will enable the
characterization of the spatial variation of the polarization
percentage, dust temperature, and dust emissivity. While the
maps will be smoothed to relatively large pixels for comparison 
with \planck, we will retain the ability
to see any fine-scale, higher-level polarization signals to help the
planning of future ground-based, balloon-borne and space experiments.

\section{INSTRUMENT}
\label{sec:inst}  % \label{} allows reference to this section

\subsection{Optical Design}
\begin{figure}
  \begin{center}
  \begin{tabular}{c}
    \includegraphics[height=5.5cm, keepaspectratio]{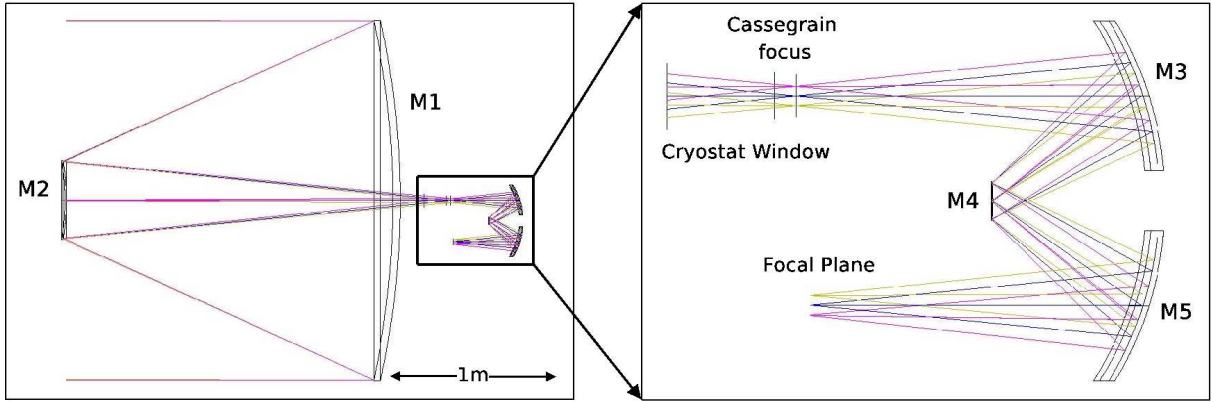}
  \end{tabular}
  \end{center}
  \caption[example] { \label{fig:optical}  Schematic of the optical layout for the
    \blast\ telescope and receiver is shown on the left, with the 1.5\,K
    optics, located within the cryostat, shown in an expanded view on
    the right.  The image of the sky formed at the input aperture is
    re-imaged onto the bolometer detector array at the focal
    plane. The mirror M4 serves as a Lyot stop, which defines the
    illumination of the primary mirror for each element of the
    bolometer detector arrays. The three wavelength bands are separated by a
    pair of dichroic beamsplitters (not shown) which are arranged in a
    direction perpendicular to the plane, between M5 and the focal
    plane.}
   \end{figure}

\blastpol\ utilizes a Cassegrain telescope consisting of a 1.8\,m parabolic primary mirror (M1) and a 40\,cm correcting secondary (M2).  The field of view of the telescope at 250 \um\ is 13.5\arcmin\,$\times$\,6.5\arcmin\ at the Cassegrain focus. This system redirects the light to a series of cryogenically cooled (1.5\,K) re-imaging optics (M3, M4, M5) arranged in an Offner-relay configuration, where M4 is a Lyot stop.  Figure \ref{fig:optical} shows a schematic of the optical path of the BLAST telescope.

Radiation emerging from M5 is split into three frequency bands by 
low-pass edge dichroic filters\cite{ade06}, which allows us to image 
the sky simultaneously at 250, 350 and 500 \um. The first dichroic filter
reflects wavelengths shorter than 300\,\um\ and transmits longer
wavelengths. This reflected light is directed on to a filter directly 
in front of the 250\,\um\ array, which reflects wavelengths
shorter than 215\,\um, and is further defined by the waveguide frequency 
cutoff at the exit of each of the feedhorns coupled to the detector array. 
For the 350 and 
500 \um\ arrays, the band is defined at the short-wavelength end 
by the transmission of a dichroic filter and at the long-wavelength end 
by the waveguide cutoff. Each band has a 30$\%$\ width. The 
frequency performance of the filters was evaluated with 
Fourier transform spectroscopy.

Although the primary mirror was recovered after the destruction of BLAST06, eventually it was decided that a new primary mirror was needed.  The surface of the new mirror has an RMS of $\sim$1.0 \um, with the overall shape of the mirror good to $\sim$10 \um.  The secondary mirror was also recovered after BLAST06, and has been reused for \blastpol\ (after resurfacing to remove some scratches).  The estimated antenna efficiency of the telescope is $>$\,80$\%$, with losses caused by both the roughness of the primary and the quality of the re-imaging optics. More information about the optical design and performance of the BLAST telescope can be found in Refs. \citenum{Olmi02} and \citenum{pascale08}.

Temperatures of the primary and secondary mirrors do not remain constant throughout the flight.  Diurnal temperature variations of $\sim$10\deg\,C have been observed on previous BLAST flights\cite{pascale08}.  These thermal variations result in changes to the radii of curvature of various optical surfaces.  To compensate, the position of the secondary mirror with respect to the primary can be changed in flight by three stepper motor actuators.  These actuators are also used to set the original tip/tilt alignment of the secondary.  Analysis of the BLAST optical system indicates that the distance between the primary and secondary mirrors must be kept to within 100 \um\ to avoid significant image degradation.

\subsection{Detectors}

The \blastpol\ focal plane consists of 149, 88 and 43 detectors at 250, 350 and 500 \um\ respectively.  The detectors consist of silicon-nitride micromesh (``spider-web'') bolometric detectors coupled with 2$f\lambda$ feedhorn arrays.  The bolometer detector array design is based on the Herschel \spire\ instrument detectors.\cite{bock98, row03}.  Detector sensitivity is limited by photon shot-noise from the telescope.  The total emissivity for the warm optics of $\sim$ 6\%\ 
is dominated by blockage from the secondary mirror and supports. 
%We estimate the optical efficiency of the cold filters and optics to be $\eta_{\rm opt} \ge 0.3$. 

The detectors are read out with an
AC-biased differential circuit. The data acquisition electronics
demodulate the detector signals to provide noise stability to low
frequencies ($<$ 30 mHz), which allows the sky to be observed in a
slowly-scanned mode. Slow scanning is preferable to a mechanical
chopper for mapping large regions of sky.  The
data are collected using a high-speed, flexible, 22-bit data
acquisition system developed at the University of Toronto. The system
can synchronously sample up to 600 channels at any rate up to 4
kHz. Each channel consists of a buffered input and a sigma-delta
analog to digital converter.  The output from 24 channels are then
processed by an Altera programmable logic device which digitally
anti-alias filters and demodulates each input. The results then are stored
to disk.

\subsection{Polarimetry}
A photo-lithographed polarizing grid has been mounted in front of
the feed-horn arrays for each bolometer detector array.  The grids
are patterned to alternate the polarization angle sampled by
90\deg\ from horn-to-horn and thus bolometer-to-bolometer along
the scan direction.  \blastpol\ will scan so that a source on the
sky passes along a row of detectors, and thus the time required to
measure one Stokes parameter ({\em Q} or {\em U}) is just equal to the
separation between bolometers divided by the scan speed. For the
250 \um\ detector array where the bolometers are separated by
45\arcsec, and assuming a typical scan speed of 0.1\,\deg\ per
second, this time would be 0.125 seconds.  This timescale is short
compared to the characteristic low frequency (1/f) noise knee for 
the detectors at 0.035\,mHz\cite{pascale08}.

\begin{figure}
  \begin{center}
  \begin{tabular}{c}
    \includegraphics[height=8cm, keepaspectratio]{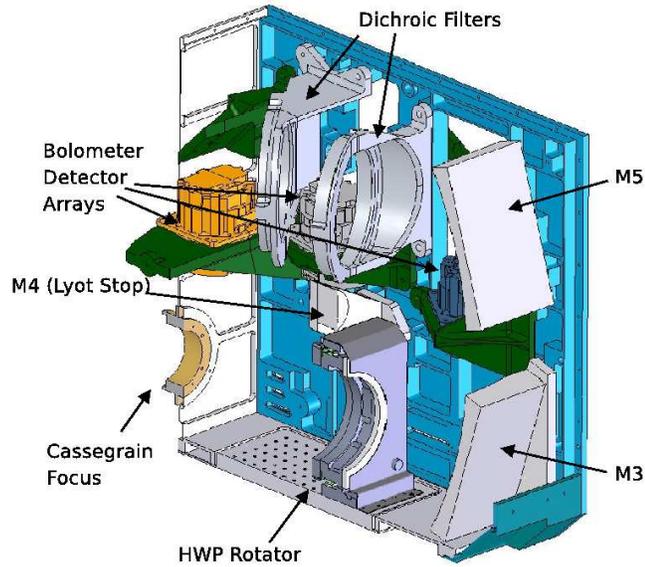}
  \end{tabular}
  \end{center}
  \caption[example]
   { \label{fig:opticsbox}
    A cutaway view of the \blastpol\ optics
box. The light enters from the lower left and is re-imaged
onto the bolometer detector arrays (BDAs). Dichroic filters split the
beam into each of the BDAs for simultaneous imaging of the sky at 
250, 350 and 500 \um. A modulating half-wave 
plate is mounted between the entrance to the optics box 
and M3, and polarizing grids have been mounted directly in 
front of each of the BDAs. The stepper motor which 
rotates the waveplate is located outside the optics box.
   }
\end{figure}

\begin{figure}
  \begin{center}
  \begin{tabular}{c}
    \includegraphics[height=5.8cm, keepaspectratio]{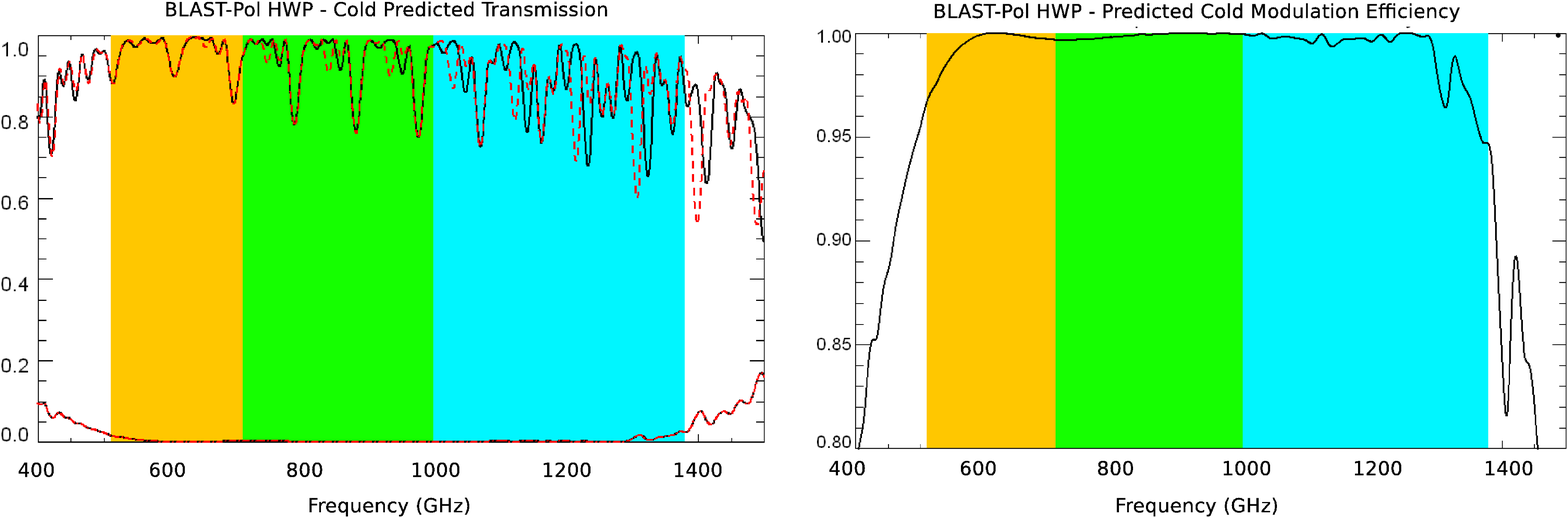}
  \end{tabular}
  \end{center}
  \caption[example]
   { \label{fig:modeff}

{\em Left:} The predicted transmissions through the HWP
as a function of frequency. The approximate extent of the BLAST 
bandpasses are also indicated. {Solid Black Curve:} transmission between 
two parallel polarizers (i.e. Q=1 to Q=1) with the HWP axis at 0 deg
 (black line).
{ Red Dashed Curve:} transmission between two parallel polarizers Q=1
 to Q=1 with the HWP axis at 90 deg (or equivalently  Q=-1 to Q=-1 
with the HWP axis at 0).
{ Red and Black Lower Curve:} transmission with the HWP axis at 0 deg between two 
perpendicular polarizers.
{\em Right:} predicted modulation efficiency
(as a function of frequency), obtained as ($T_{copol}-T_{xpol}$)/($T_{copol}+T_{xpol}$).  Note that the y-axis scale ranges from 0.8 to 1.
   }
\end{figure}

In order to measure the other Stokes parameter and provide
polarization modulation an Achromatic Half Wave Plate (AHWP) has
been incorporated into the optical design as shown in Figure
\ref{fig:opticsbox}.  The HWP is mounted inside the optics box
19.1\,cm from the Cassegrain focus of the telescope to minimize
the modulation of any potential plate local defects.  The predicted modulation 
efficiency across the \blast\
bands is given in Figure\ \ref{fig:modeff}.

The use of a rotating HWP as a polarization modulator is a
widespread technique at millimeter and submillimeter wavelengths
(see for example Refs. \citenum{EBEX}, \citenum{Savini}, \citenum{Pisano} and
\citenum{MAXIPOL}). The \blastpol\ HWP is 
10\,cm in diameter and is
constructed from 5 layers of sapphire, each 500\um\,in thickness.
The layers are glued together with a 6\,\um\ layer of
polyethylene, and an anti-reflection coating (made from metal mesh
filter technology, see Ref. \citenum{Jin}) is glued to each surface of the half-wave plate. Figure \ref{fig:modeff} shows the predicted transmissions and modulation efficiencies for the BLAST HWP as a function of frequency.  They are based on a comprehensive set
of data taken at room temperature, which has been extrapolated to 
estimate the 4\,K performance of the HWP.

We operate the HWP in a stepped mode, rather than a continuously 
rotating mode.  The rotator employs a pair of thin-section steel 
ball bearings housed in a stainless steel structure, and is driven 
via a gear train and a G-10 shaft leading to a stepper motor 
outside the cryostat.  A ferrofluidic vacuum seal is used for the 
drive shaft.  The angle sensing at liquid Helium temperatures is 
accomplished by a potentiometer element making light contact with 
phosphor bronze leaf springs.  During operation, we will carry 
out spatial scans at eight HWP angles spanning 180 degrees of 
rotation (22.5\deg\ steps).  The rotator and encoder are based 
on a design used successfully at South Pole\cite{renb04}.

\subsection{Cryogenics}
The receiver consists of an optical cavity inside a long hold-time
liquid-nitrogen and liquid-helium cryostat. Both the nitrogen and
helium are maintained at slightly more than atmospheric pressure
during the flight to minimize loss due to pressure drop at altitude.
A \3he refrigerator maintains the detectors at 300\,mK during
flight. The self-contained, recycling refrigerator can maintain a base
temperature of 280\,mK with 30\,$\mu$W of cooling power for 4 days. It
can be recycled within 2\,hr. The \3he refrigerator uses a pumped \4he
pot at $\sim 1$\,K for cycling and to increase the hold time of the
system. The pumped pot maintains 1\,K with 20\,mW of cooling power
with outside pressure 15 Torr or less. The entire optics box
containing the re-imaging optics is also cooled to 1 K.

\subsection{Gondola}
\begin{figure}
  \begin{center}
  \begin{tabular}{c}
    \includegraphics[height=7cm, keepaspectratio]{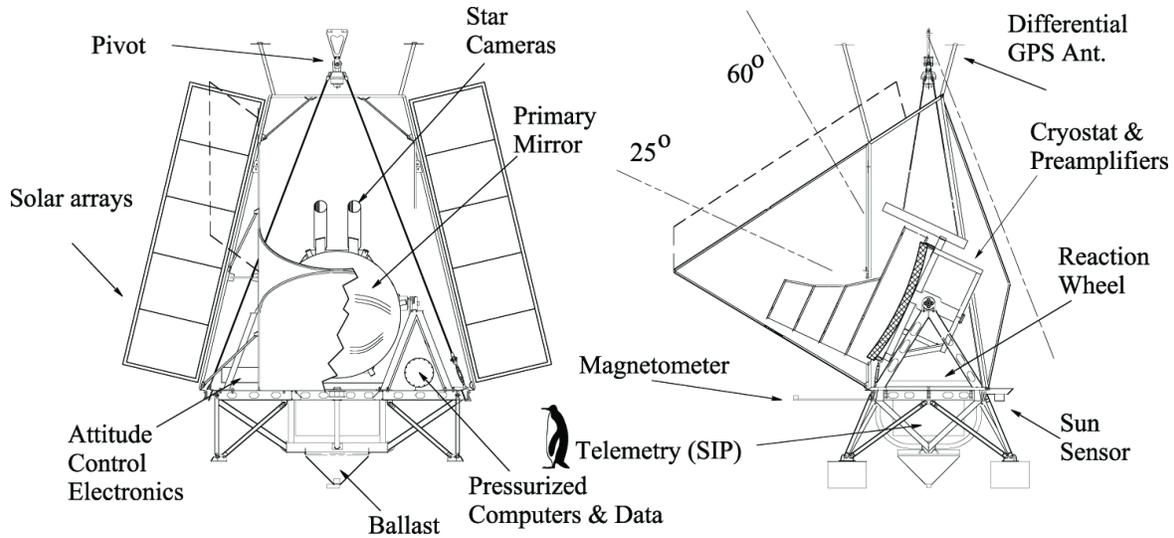} 
  \end{tabular}
  \end{center}
  \caption[example]
   { \label{fig:blastlayout}
 Front and side schematic drawings of the \blast\ gondola. 
A 1-m tall Emperor penguin is shown for scale.  The inner 
frame, which can be pointed in elevation, consists of the 
two star cameras, the telescope and its light baffle, the 
receiver cryostat, and associated electronics. The telescope 
baffles and sunshields have been updated for \blastpol. 
}
\end{figure}

The \blastpol\ gondola provides a pointed platform for the telescope 
and attaches to the balloon flight train.  The gondola consists of 
two parts: an outer aluminum frame, which can be pointed in azimuth; 
and an inner aluminum frame which points in elevation.  Figure 
\ref{fig:blastlayout} shows a schematic layout of the BLAST gondola 
with several features labelled. 

\begin{figure}
  \begin{center}
  \begin{tabular}{cc}
  \includegraphics[height=7cm, keepaspectratio]{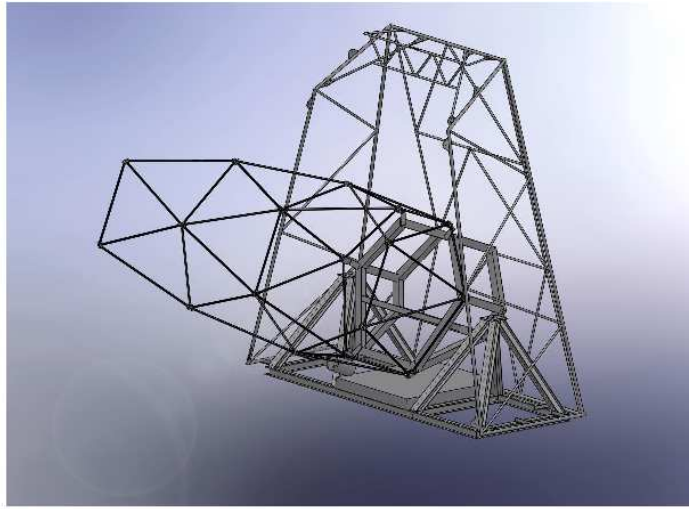}
  \end{tabular}
  \end{center}
  \caption[example]
   { \label{fig:frames}
A drawing of the \blastpol\ gondola showing just the inner 
and outer frame gondola structures, including 
the new inner frame sun shields which will allow the telescope to point 
to an azimuth distance of 45\deg from the Sun.
}
\end{figure}

The outer frame is a suspended from a $1.1 \times
10^6$\,m$^3$ helium balloon, provided by NASA's Columbia 
Scientific Balloon Facility, through a steel cable ladder 
and parachute.  Control systems, including flight computers
 and telemetry systems are mounted on the outer frame.  
Data are stored on solid state disks on the computers.  
Some portion of the data can be transmitted to a ground 
station by satellite links. The inner frame houses the 
mirrors, the receiver, the receiver read-out electronics and 
primary pointing sensors. These are all rigidly mounted with
respect to each other on the inner frame in order to ensure 
that mechanical alignment is maintained throughout the flight.

To avoid large thermal changes in the optics both the inner 
and outer frames have attached sunshield structures.  Figure 
\ref{fig:frames} shows the \blastpol\ sunshields.  Shields 
on the outer frame are constructed from aluminized mylar and 
mounted on an aluminum frame, and are similar to those used 
in previous \blast\ flights.  In addition, for \blastpol\ we 
have constructed new shields which are attached to a carbon 
fiber frame and are mounted to the inner frame.  This 4-m shield 
will allow us to point the telescope to within 45\deg\ of the Sun, 
in order to observe targets close to the Galactic Center.  

Telescope pointing is controlled by three motors.  The azimuth 
pointing is controlled by a brushless, direct drive servo motor 
attached to a high moment of inertia reaction wheel, and an 
active pivot motor which connects the cable suspended gondola 
to the balloon flight train.  The reaction wheel consists of a 1.5-m
disk made of 7.6\,cm thick aluminum honeycomb, with 48 0.9\,kg 
brass disks mounted around the perimeter.  The reaction wheel 
is mounted at the center of mass of the telescope, directly 
beneath the active pivot.  By spinning the reaction wheel 
angular momentum can be transferred to and from the gondola,  
allowing precise control over the azimuth velocity 
of the telescope with minimal latency. The active pivot motor 
provides additional azimuthal torque 
by twisting the flight train, and can also be used over long 
time scales to transfer angular momentum to the balloon.

The elevation of the inner frame is controlled by a servo motor 
mounted on one side of the inner frame at the attachment point 
to the outer frame. A free bearing provides the connection point 
between the inner and outer frames, on the other side.

In-flight pointing is measured to an accuracy of $\sim$30\arcsec\ 
by a number of fine and coarse pointing sensors.  These include 
fiber optic gyroscopes, optical star cameras, a differential GPS, 
an elevation encoder, inclinometers, a magnetometer and a Sun 
sensor. Post-flight pointing reconstruction uses only the
gyroscopes and day-time star camera\cite{rex06}. The algorithm is
based on a similar multiplicative extended Kalman
filter\cite{markley03} technique used by \wmap\cite{harman05}, which has been modified\cite{pittelkau01} to allow for the evaluation of the star
camera and gyroscope alignment parameters. The offset between the
star cameras and the submm telescope will be measured by repeated 
observations of pointing calibrators throughout the flight. For the 
BLAST06 flight it was found that the relative pointing between the 
star cameras and telescope varied as a function of the inner frame elevation, 
requiring elevation dependent corrections to pitch and yaw with scales of
$\sim$125\arcsec\ and $\sim$20\arcsec, respectively.  Post flight 
absolute pointing accuracy for BLAST06 was found to be $< 2\arcsec$, 
and random pointing errors were found to be $< 3\arcsec$ 
rms\cite{pascale08,spie08}.

%\begin{table}[h]
%\caption{Properties of the BLAST-pol bands including expected sensitivities.} 
%\label{tab:bands}
%\begin{center}       
%\begin{tabular}{|l|l|l|l|} %% this creates two columns
%\hline
%\rule[-1ex]{0pt}{3.5ex}  Central wavelength (\um)   & 250 & 350 & 500  \\
%\hline
%\rule[-1ex]{0pt}{3.5ex}  Number of Detectors        & 149 & 88  & 43  \\ %280 detectors
%\rule[-1ex]{0pt}{3.5ex}  FWHM (\arcsec)             & 30  & 41  & 59  \\ 
%\rule[-1ex]{0pt}{3.5ex}  Background power (pW)      & 50  & 36  & 26  \\
%\rule[-1ex]{0pt}{3.5ex}  Background power ($\times 10^{-17}\,W/\sqrt{Hz}$)
%                                                    & 20  & 14  & 10  \\
%\rule[-1ex]{0pt}{3.5ex}  NEFD (mJy\,s$^{0.5}$)         & 236  & 241  & 239  \\
%\rule[-1ex]{0pt}{3.5ex}  $\Delta S$(1$\sigma$,1\,hr)(1 sq. deg.) \ \ (mJy)                                                          & 38  & 36  & 36  \\
%\hline
%\end{tabular}
%\end{center}
%\end{table} 

%%%%%%%%%%%%%%%%%%%%%%%%%%%%%%%%%%%%%%%%%%%%%%%%%%%%%%%%%%%%%
\acknowledgments     %>>>> equivalent to \section*{ACKNOWLEDGMENTS}       
 
The \blast\ collaboration acknowledges the support of NASA through
grant numbers NAG5-12785, NAG5-13301 and NNGO-6GI11G, the Canadian
Space Agency (CSA), the Science and Technology Facilities Council
(STFC), Canada's Natural Sciences and Engineering Research Council
(NSERC), the Canada Foundation for Innovation, the Ontario Innovation
Trust, the Puerto Rico Space Grant Consortium, the Fondo Istitucional
para la Investigacion of the University of Puerto Rico, and the
National Science Foundation Office of Polar Programs;
C.~B. Netterfield also acknowledges support from the Canadian
Institute for Advanced Research.  L.~Olmi would like to acknowledge
Pietro Bolli for his help with Physical Optics simulations during the
testing phase of BLAST06.  We would also like to
thank the Columbia Scientific Balloon Facility (CSBF) staff for their
outstanding work, Dan Swetz for
buliding the Fourier transform spectrometer, and Luke Bruneaux, Kyle
Lepage, Danica Marsden, Vjera Miovic, and James Watt for their
contribution to the project.

%%%%%%%%%%%%%%%%%%%%%%%%%%%%%%%%%%%%%%%%%%%%%%%%%%%%%%%%%%%%%
%%%%% References %%%%%

\bibliography{refs}   %>>>> bibliography data in report.bib
\bibliographystyle{spiebib}   %>>>> makes bibtex use spiebib.bst

\end{document}